%% file: main.tex
\documentclass{article}
\usepackage[left=2cm, right=2cm]{geometry}
\usepackage[utf8]{inputenc}
\usepackage{graphicx}
\usepackage{comment}
\usepackage{caption}
\usepackage[version=4]{mhchem}
\usepackage{float}
\usepackage{url} 
\usepackage{tikz}
\usepackage{amsfonts}
\usepackage{tikz-3dplot,pgfplots}
\usepackage{pst-3dplot}
\usepgfplotslibrary{fillbetween}
\pgfplotsset{compat=1.14}
\usepackage{authblk}

\title{The space environment particle density in Low Earth Orbit based on two decades of in-situ observation}

\author[1,*]{Soumaya Azzi}
\author[2]{Xanthi Oikonomidou}
\author[3]{Stijn Lemmens}

\affil[1]{Space Debris Office, European Space Agency (ESA/ESOC), Robert-Bosch-Str. 5, 64293 Darmstadt, Germany}
\affil[2]{GMV @ Space Debris Office, European Space Agency (ESA/ESOC), Robert-Bosch-Str. 5, 64293 Darmstadt, Germany}
\affil[3]{Space Debris Office, European Space Agency (ESA/ESTEC), Keplerlaan 1, AZ Noordwijk 2201, the Netherlands}
\date{}                     
\begin{document}
\maketitle

\section{Abstract}
Currently the only method to establish the prevalence of particles, space debris or meteoroids, sized between 1 micrometre and a few centimetres, in Earth orbit is by instruments or witness plates dedicated to in-situ detection. 
Derived usable datasets are remarkably scarce and generally only cover a short period of time and a single orbital region.
Nonetheless, space environment models use those limited datasets as anchor points to extrapolate results to the entirety of Earth orbit, from the beginning of the space age to decades into the future.
Here we present a readout of over 20 years of DEBris In orbit Evaluator 1 (DEBIE-1), an in-situ detector that was launched in October 2001, providing the longest continuous set of measurements available to date.
The dataset has not been used in the generation of space environment models and hence provides a first independent source for the detection of environment changing events and for the calibration of long term evolution models.
\section{Introduction} 
The launch of Sputnik 1 on October 4, 1957 marked the beginning of the space age as it is known today. 
In the decades that followed, the number of human-made objects in space has grown exponentially. 
In addition to the derelict human-made objects left in orbit and their fragments, i.e., space debris, other sources of particles in the space environment include objects from explosions and collisions, \ce{Al2O3} dust particles and slag fragments from solid rocker motor firings, sodium-potassium alloy (NaK) droplets from reactor core ejections of nuclear-powered satellites, surface degradation particles commonly denoted as paint flakes, or ejecta released from the hypervelocity impacts in space.

As of today, 32680 space debris objects are tracked regularly by ground Space Surveillance Networks \cite{ESAcount}. 
These comprise the known space debris population, for which a catalogue is maintained corresponding roughly to centimetre sized objects in Low Earth orbit (LEO) (orbits with a geodetic height below 2000 km) and above half a metre in Geostationary Orbit (an orbit at 35786 km above the Earth equator). 

Encounters with catalogued objects can be predicted and when necessary avoided with collision avoidance manoeuvres. 
The majority of particles, however, cannot be tracked and catalogued, and hence their prevalence has to be estimated by statistical models such as the Meteoroid and Space Debris Terrestrial Environment Reference model (MASTER) of the European Space Agency (see Methods Section \ref{sec:Master}).
The model estimates that on November 2016 there were 36500 space debris objects with size greater than 10 cm, 1000000 objects between 1 cm and 10 cm, and 130 million objects between 1 mm and 1 cm. In addition to space debris particles, Earth orbits contain a natural meteoroid environment dominated by particles in the millimetre and sub-millimetre size regime.
 
While human-made space debris and its modelling was not a globally recognised concept in the 1960s, the natural meteoroid environment and the effect it might have on spacecraft (i.e., surface degradation, danger for astronauts etc.) was already of great interest. 
First, NASA launched its Pegasus project in 1965 with the objective of studying the frequency of small meteoroid impacts on spacecraft. 
In the following decades, and with the increase of human-made objects in space modelling efforts have increased, with returned space-exposed hardware and surfaces providing screenshots of the small-sized space debris environment in low orbits. 

\begin{figure}
\centering
\includegraphics[scale=0.85]{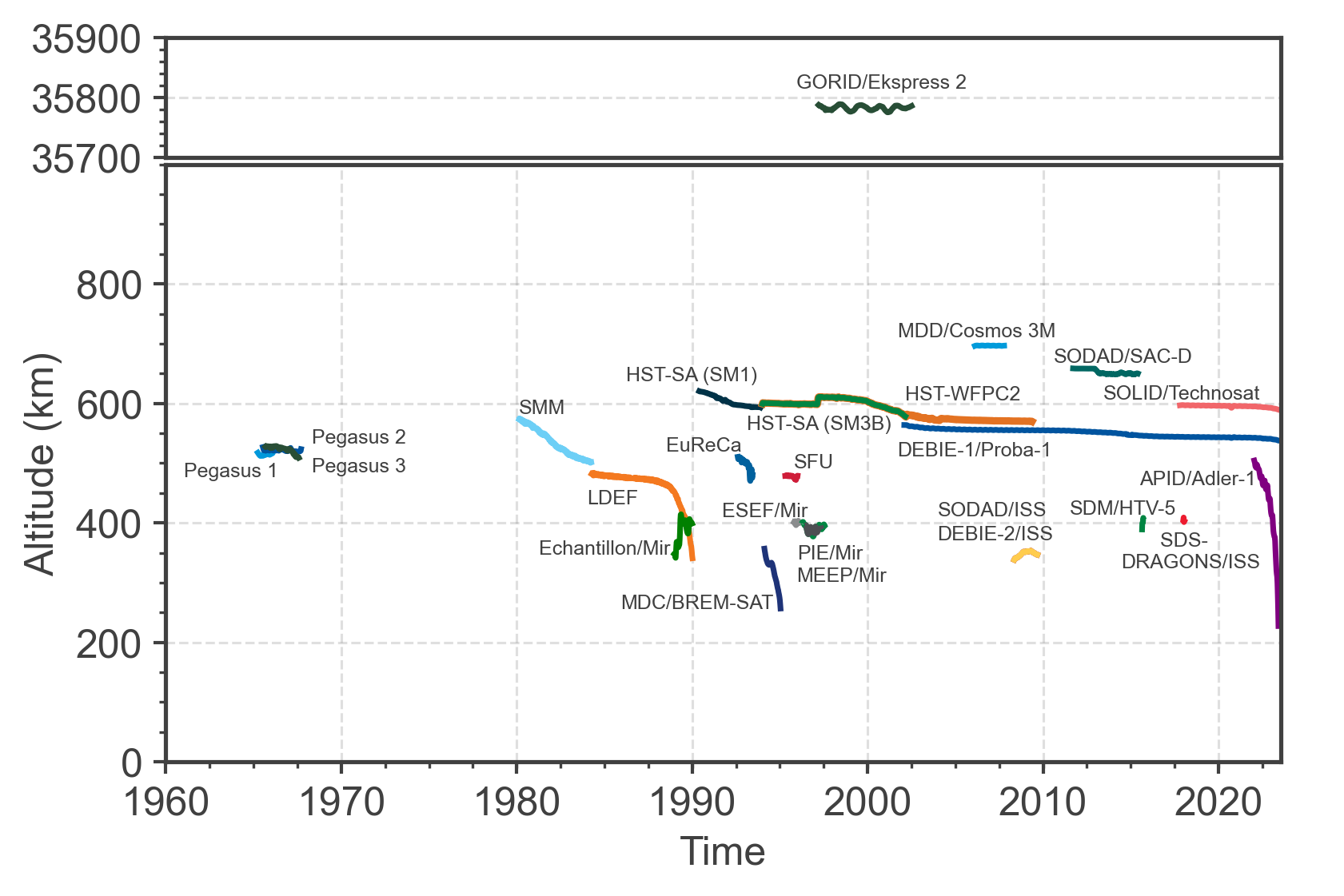}
\caption{In-situ meteoroid and space debris detectors in orbit since 1960. The plot includes surfaces and experiments as well as active in-situ detectors that flew on-board spacecraft, launchers or space stations in LEO and GEO between 1960 and mid-2023. Sensors and experiments in lunar or inter-planetary orbits as well as sensors for which their operation has not been verified are not included (list not exhaustive). For the representation of the orbital altitude, altitude at perigee is assumed.}
\label{fig:in_situ_sensors}
\end{figure}

Apart from spacecraft and surfaces returned to Earth, active in-situ detectors can additionally provide time-stamped measurements. 
While various experiments and sensors have flown onboard platforms over the past three decades (Figure \ref{fig:in_situ_sensors}), very few sensors have provided data that is statistically meaningful and of sufficient quality to be used by space environment models.
This is largely due to the high sensitivity of most active sensors to surrounding noise. The resulting data requires sophisticated post-processing. In many cases, and due to the design complexity of such sensors, malfunctions in their system have rendered the resulting data, if any, unusable. 
At present, only four sources have been used by MASTER for its small particle validation: the measurement data from Long Duration Exposure Facility (LDEF), the solar panels retrieved during the first and third Hubble Space Telescope Service Missions (HST-SM1 and HST-SM3B) and the European Retrievable Carrier (EuReCa). 

As of today, the Debris In-Orbit Evaluator (DEBIE-1) is the only detector which has been providing measurements in LEO for more than two decades. 
Data from the detector has been analysed and post-processed, making it the first in-situ detector candidate whose measurements could be used for the validation chain of the MASTER model and the updates of its population. 
Incorporating this data set into the MASTER calibration process is outside the scope of this work. Instead, the focus is on conducting an exploratory statistical analysis of the collected data and examining the trends it reveals.

The Debris In-Orbit Evaluator (DEBIE) detector was devised in 1996 as a low cost and low resource add-on dust sensor for spacecraft. 
The instrument is equipped with two impact plasma detectors (ion and electron), two piezoelectric momentum detecting sensors, and a thin (6 {\textmu}m) aluminium foil to gather penetration statistics \cite{J_S_PhD}. 
Its operational concept is based on measuring the plasma generated by the particle impacts on the foil. 
The two piezoelectric transducers are coupled mechanically to the foil and measure the momentum of the impact. 
Particles with sufficient energy to penetrate the foil are detected by the plasma detector (electrons) placed behind the foil. 
The particle speed and mass can be calculated from the measured parameters with the aid of predefined calibration parameters derived from on-ground calibration campaigns \cite{kuitunen2001debie,J_S_PhD}.

The detector was originally planned to collect data from three different orbits, with DEBIE-0 on STRV-1c sent to a Geostationary Transfer Orbit (GTO), DEBIE-1 on PROBA-1 to a polar orbit at around 600 km, and DEBIE-2 on the International Space Station to a lower LEO. 
Due to the receiver failure of STRV-1c, no data from DEBIE-0 has became available \cite{J_S_PhD}. However, both DEBIE-1 and DEBIE-2 collected data from their corresponding orbits.

PROBA-1 is a technology demonstration microsatellite launched on 22 October 2001 that is still in operation today (end 2023). 
DEBIE-1 consists of two DEBIE Sensor Units (SU-1 and SU-2), with one facing the flight direction and the other the starboard side  (right-hand side with respect to SU-1). 
The detection area of each sensor unit is $10\times10$ cm$^2$. 

The instrument was initially calibrated to detect sub-millimetre particles with masses down to $10^{-15}$ g depending on the impact speed.  
The threshold diameters (smallest particles detected in space) for SU-1 and SU-2 have been estimated in \cite{J_S_PhD} using  data from the first few years of DEBIE-1 and correspond to 1.52 and 3.03 micrometres respectively (assuming an average density of 2.5 g/cm³). 

The raw data from DEBIE-1 was processed and filtered to remove noise in the dataset \cite{bunte2021data,DMF04}. Evaluating the effectiveness of the filtering process is outside the scope of this paper. Additionally, \cite{J_S_PhD} and \cite{DMF04} indicate that the speed data is unreliable due to substantial noise-related uncertainties, which would lead to highly uncertain mass values. Therefore, only time-stamped impact counts were used in this study.

In this paper, the impacts from DEBIE-1 between 2002 and 2022 are collected, filtered and compared to available simulation tools. 
Particular trends in the obtained measurements are examined, and possible correlations with major fragmentation events that took place around the orbit of PROBA-1 are studied. Section \ref{sec:Results} describes the main results from the analyses, followed by a discussion addressing the added value to the modelling of the space debris density of sub-millimetre objects in LEO. 
The encountered limitations and the necessary focus aspects for future efforts on statistical modelling of space debris and meteoroids using active in-situ sensors are additionally discussed.

\section{Results}\label{sec:Results}
\begin{figure}
\noindent\begin{minipage}{.5\textwidth}
\centering
\includegraphics[width=0.9\textwidth]{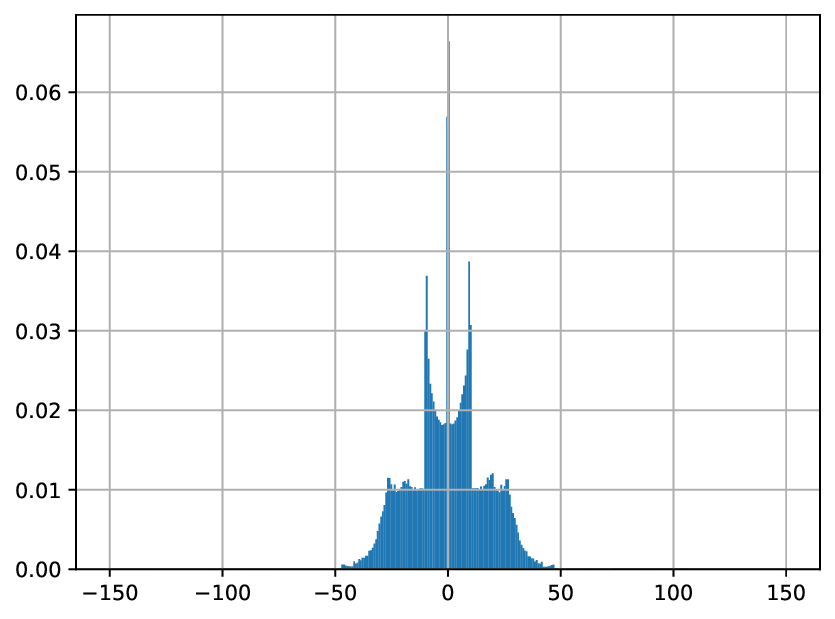}            
\end{minipage}
\begin{minipage}{.5\textwidth}
\centering
\includegraphics[width=0.9\textwidth]{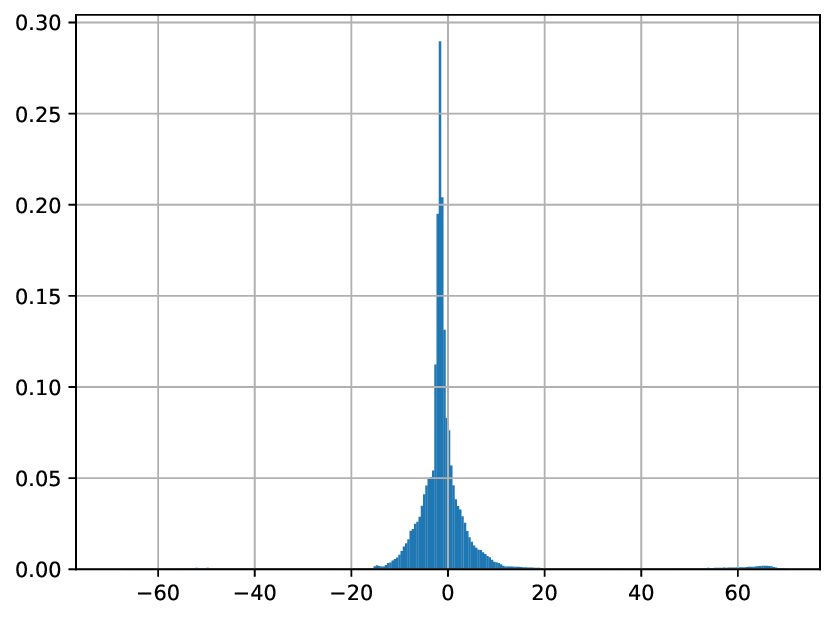}          
\end{minipage}
\caption{Probability density function of azimuth (left) and elevation (right) in degrees, from 2002 to 2022.}
\label{fig:AziElevPDF}
\end{figure}

At this stage, the detector surface is modelled in MASTER (see Methods Section \ref{sec:Master}) as an orbiting target. 
The exact orbit of PROBA-1 is retrieved from the housekeeping telemetry and used to simulate the flux from  both natural and human-made impact sources. 
The attitude of PROBA-1 was also taken into account by computing the azimuth and elevation of the detector in the orbit frame.
Figure \ref{fig:AziElevPDF} shows the distribution of the azimuth and elevation angles of DEBIE-1 during the last $20$ years, forming the basis for assuming that the SU-1 sensor faces, on average, the  flight direction. 
MASTER consistently provides the mean flux throughout a single orbital revolution, accommodating slight alterations in  spacecraft  attitude  through averaging.

\begin{figure}
\centering
\includegraphics[scale=0.4]{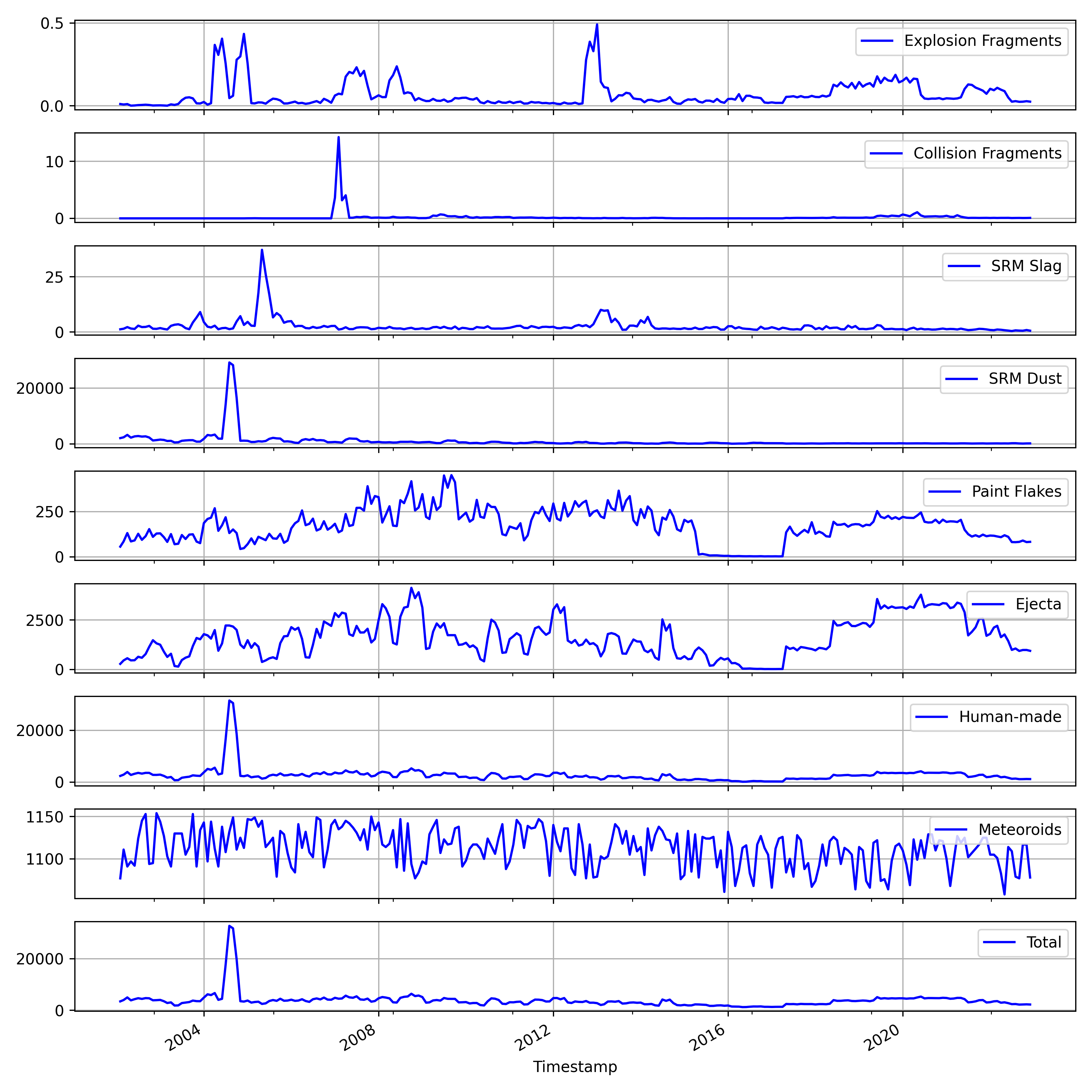}
\caption{The flux, estimated by MASTER, measured in $1/\text{m}^2/year$ from different particle sources, arranged from top to bottom: explosion fragments, collision fragments, dust and slag from Solid Rocket Motor (SRM) firings, surface degradation particles (paint flakes), ejecta, aggregated into human-made debris, then meteoroids, cumulatively representing the  total flux, as extracted from MASTER simulating DEBIE-1, over 2002-2022.}
\label{fig:DiffSourcesMaster}
\end{figure}

From MASTER a yearly expected flux value is generated monthly over 2002-2022.
Figure \ref{fig:DiffSourcesMaster} shows the breakdown of the flux from different particle sources. 
The Gr\"{u}n model was used here to model the meteoroid environment with the Taylor velocity distribution \cite{grun2012interplanetary}. 
According to the model predictions, the major contributions to the total simulated flux come from ejecta (i.e. particles generated following a hypervelocity impact) and meteoroids.
Late 2004, a notable surge is observed in the estimated SRM dust population which stands out from the predicted background environment.

\begin{figure}
\captionsetup{justification=justified}  
\begin{minipage}[b]{.5\textwidth}  
  \centering  
  \includegraphics[scale=0.47]{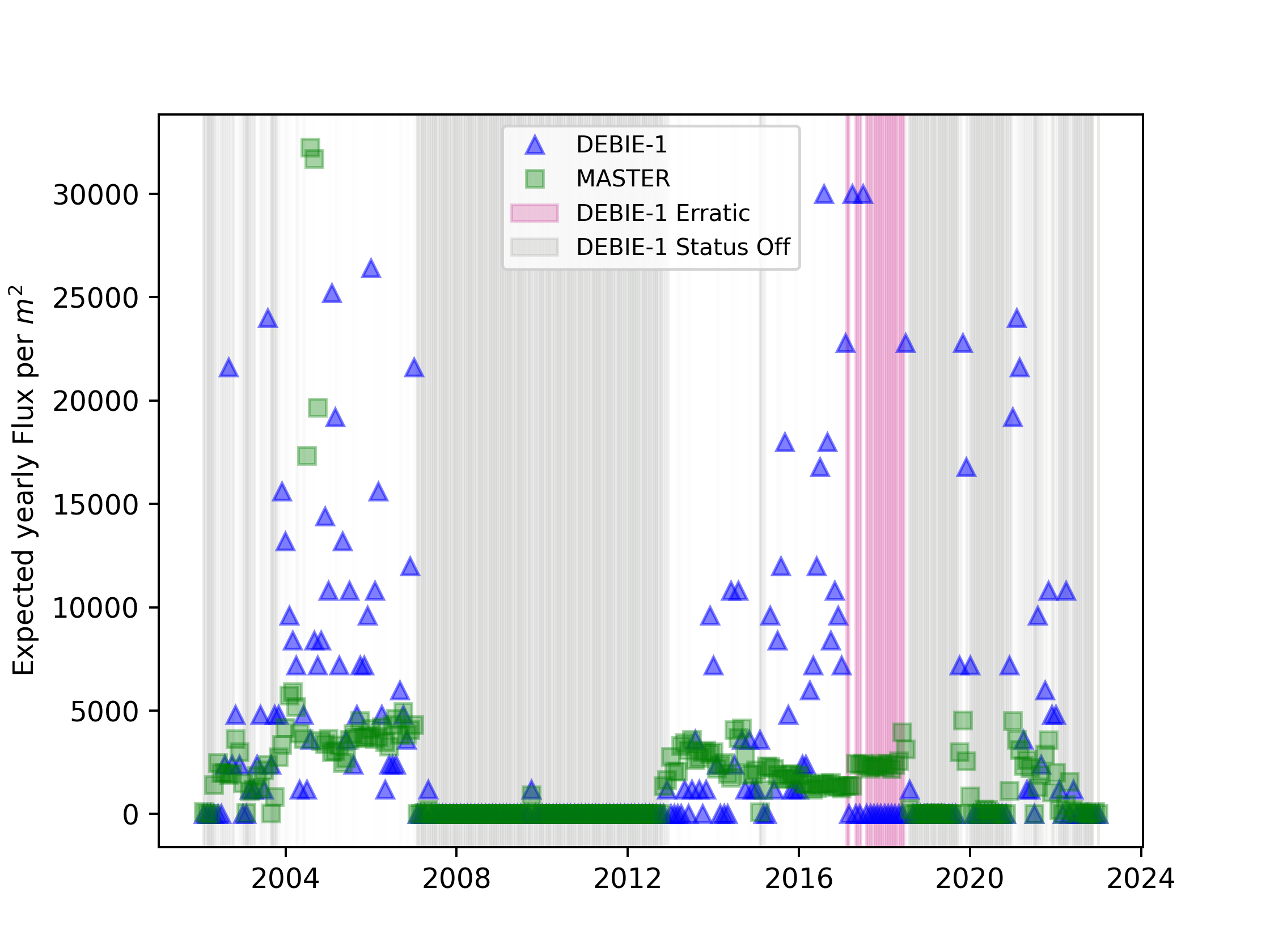} 
\end{minipage}%
\begin{minipage}[b]{0.5\textwidth}  
  \centering  
  \includegraphics[scale=0.47]{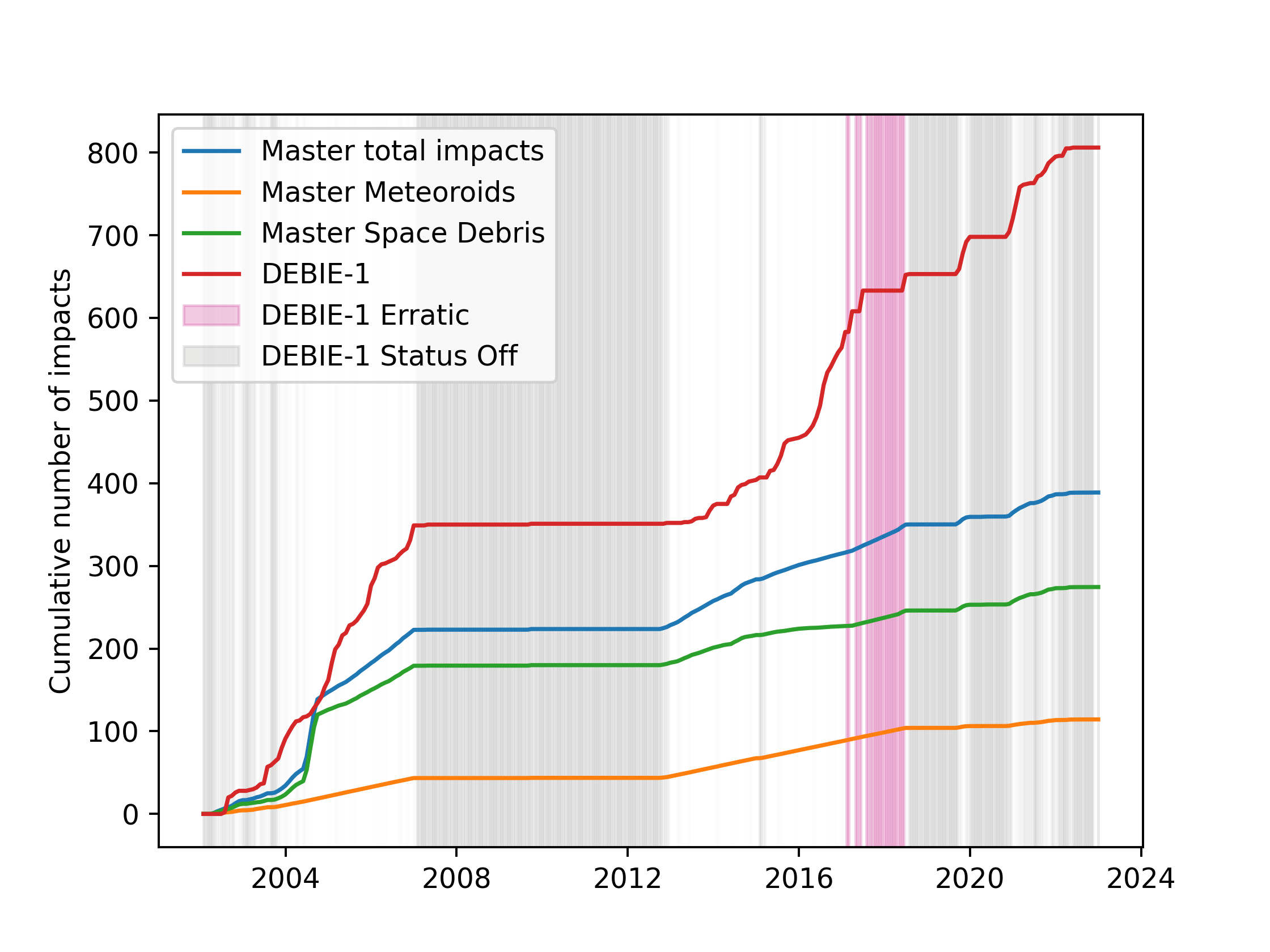}  
\end{minipage}
\par
\begin{minipage}[t]{.45\textwidth}
  \caption{In blue markers, the expected yearly flux on a 1 $\text{m}^2$ surface simulated monthly with MASTER. In green markers, the yearly flux per 1 $\text{m}^2$ surface detected in a month by DEBIE-1. DEBIE-1 status in the background, grey when the sensor was switched off most of the month, purple when the data from the sensor was erratic. }
  \label{fig:Debie_Master}  
\end{minipage}
\hfill
\begin{minipage}[t]{.45\textwidth}  
  \caption{Cumulative number of impacts from space debris and meteoroids, simulated with MASTER versus measured by DEBIE-1. DEBIE-1 status in the background, grey when the sensor was switched off most of the month, purple when the data from the sensor was erratic. } 
  \label{fig:Cum_impacts}
\end{minipage}  
\end{figure}

The particle flux is then normalised proportionally to when DEBIE-1 was switched on to compare the model estimates and the detector readings. Prior to that, the data from DEBIE-1 was processed using noise filtering techniques described in Section \ref{sec:data_availability}. 
Figure \ref{fig:Debie_Master} demonstrates a comparison over time between the total flux from MASTER and the corresponding flux from DEBIE-1.
Figure \ref{fig:Cum_impacts} shows a comparison in cumulative impacts count between the sensor and space debris and meteoroids as simulated by MASTER.
One can notice that the flux and the impact counts from DEBIE-1 and MASTER are within the same order of magnitude, and exhibit similar patterns. Both the model and the sensor show an upward trend in detection rates around 2005. In MASTER, this rise is attributed to the SRM firing contribution, as shown in Figure \ref{fig:DiffSourcesMaster}. On the other hand, DEBIE-1 potentially detected the same dust, which could explain the concurrent increase in detections.

\begin{figure}
\centering
\includegraphics[scale=0.7]{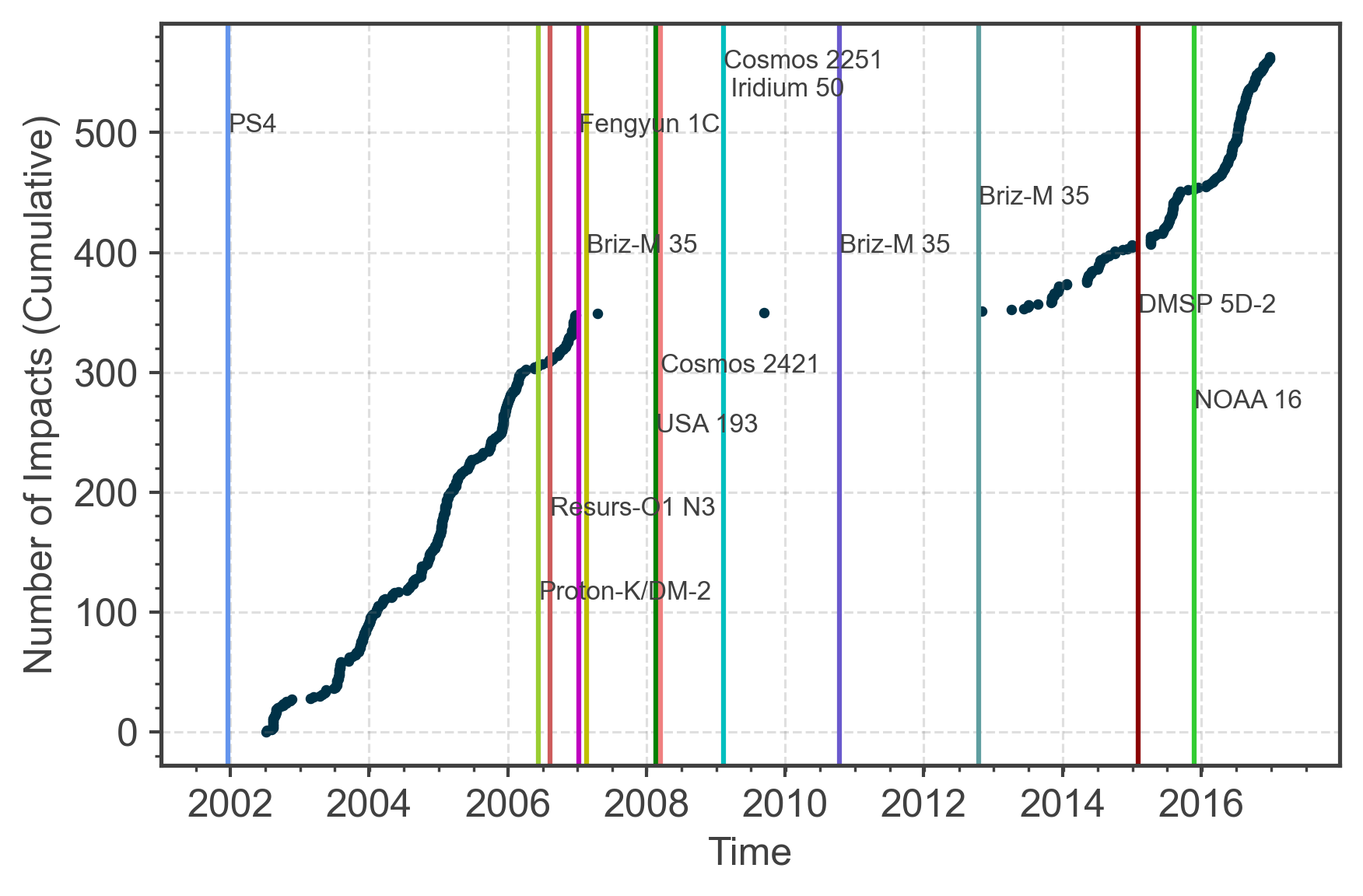}
\caption{Cumulative number of impacts captured by SU-1 over time against major fragmentation events (catalogued fragments $\geq$ 100) in LEO. The number of impacts is computed from the filtered data of SU-1 between 2002 and 2016.}
  \label{fig:impactsVSevents}
\end{figure}

A second increase in the detection rate of DEBIE-1 can be observed in 2015-2016, in accordance with the increase in cumulative impact counts from Figure \ref{fig:impactsVSevents}. 
We thus see a deviation of the MASTER model, which can be due to an underestimation in the human-made and/or the natural particle flux levels of the model. 
The difference could also be attributed to DEBIE-1 being sensitive to particles smaller than 1 {\textmu}m, or to nonfiltered false positive impacts remaining in the dataset.

To confirm potential underestimations in the model, especially during the period from 2001 to 2016 when calibration of the MASTER model occurred using observed data from space surveillance and witness plates like LDEF, HST, and EuReCa, a direct comparison based on impact counts can be conducted.

An analysis of the impacts detected by DEBIE-1 to major fragmentation events in its orbital vicinity shows that the number of impacts rose shortly after two main events and a possible correlation can be assessed.
Figure \ref{fig:impactsVSevents} shows an increase in number of impacts suspected to be related to the break-ups of DMSP 5D-2 (1995-015A) on 3 February 2015 and NOAA 16 (2000-055A) on 25 November 2015 \cite{NASA_2023history}. 
One can compare the empirical yearly mean of the detected impacts before and after the events. Through 2013 and 2014 a mean of 29.5 impacts per year is assessed. 
During and after the fragmentation events, so 2015 and 2016, the mean value of impacts per year rises to 80. 
The number of detected impacts by the sensor can be modelled by a Poisson distribution $X \sim \mathit{Pois}(\lambda)$, 
 and the rate of the Poisson distribution would be then $\hat{\lambda} = 29.5$, the maximum likelihood estimator. 
In assessing a potential increase in impacts, our evaluation of $ \mathbb{P} (X \leq 80) = 0.99$ leads to the conclusion that, at a $2\%$ significance level, there has been a significant rise in the number of impacts during 2015 and 2016.

Comparing these results to a simulated oriented surface in a similar orbit as DEBIE-1 using the modelling tool MIDAS (see Section \ref{sec:midas} in Methods for more details), an increase is not detected, as a matter of fact the yearly simulated total impacts stayed roughly constant with 30, 24, 22 and 15 yearly impacts throughout 2013-2016.

To further investigate the initial hypothesis of a potential causality between the 2015-2016 increase in impacts and  fragmentation events, the DMSP  5D-2 explosion is simulated using POEM (Section \ref{sec:poem} in Methods for more details).
POEM models the explosion and generates detailed fragment clouds as well as their orbits and evolution over time.

By considering a volume in space where fragments susceptible to cross PROBA-1 would orbit, one can deduce the density of fragments from the simulated explosion. 
The densities are computed by dividing the count of fragments over a volume in space, in counts per $\text{m}^3$. 
The aim is to compare the simulated density to the one from the sensor, in the respective volume of space where the simulated fragments and PROBA-1 orbit. 
The density derived from the simulated event is calculated within a volume defined by the  right ascension of the ascending node (RAAN) boundaries of the fragments over time, extending 10 km above and below PROBA-1 orbit.  
The specific details outlining volumes and densities computations are provided in Section \ref{density definition}.

\begin{figure}
\centering
\includegraphics[scale=0.6]{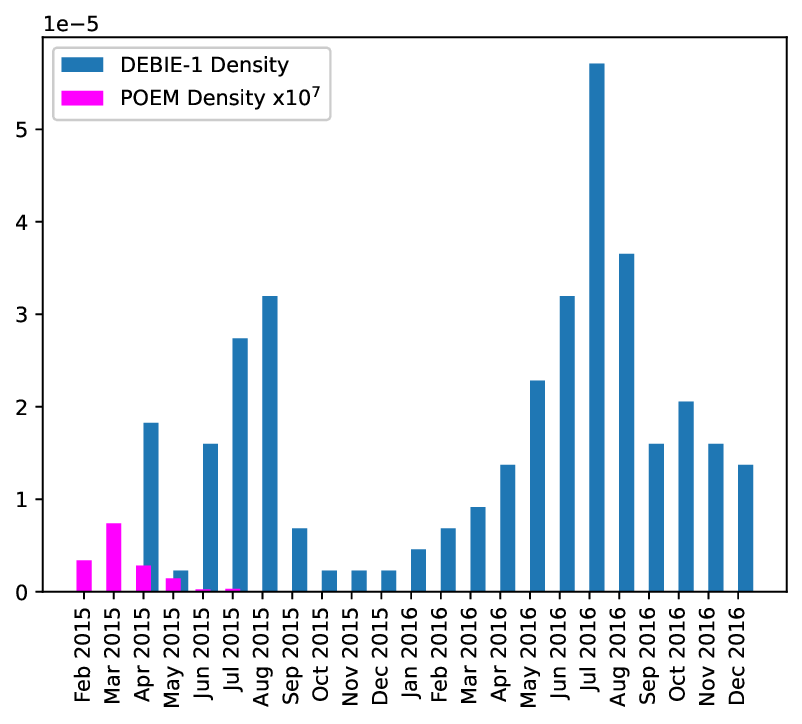}
\caption{Monthly debris density, comparison between POEM and DEBIE-1 in $\frac{counts}{\text{m}^3}$}
\label{fig:poemVSdebie}
\end{figure}

Figure \ref{fig:poemVSdebie} summarises the monthly density comparison between POEM and the detected impacts by DEBIE-1, for approximately 2 years  after the explosion of DMSP  5D-2. 
The sensor was off $40\%$ of the time in February 2015 but turned on till December 2016. 
No impacts were registered during February and March 2015. 
The densities differ significantly, with approximately 7 orders of magnitude between the sensor and the cloud model. 
Four to five months after the explosion, the simulated cloud of fragments starts diverging making the affected volume grow larger.
Additionally, the number of fragments related to the explosion decreases with time, as fragments escape the predefined volume because of the RAAN drift due to the Earth's J2 perturbation \cite{klinkrad2006space}, resulting in notably low density.
The impacts on DEBIE-1, however, seem to peak in summer 2015 and 2016 so no direct correlation with the DMSP 5D-2 explosion can be concluded, at least not from fragments that are in DEBIE-1's detection range.

The analysis concludes that while some SRM dust flux can be correlated to the rise in the DEBIE-1 detection in the early operations, space debris created in fragmentation events does not correlate with the DEBIE-1 data. 
Rather the sensor seems to sample ejecta, SRM dust and meteoroids. \cite{DMF04TN} (Figure 3) also shows no clear latitude-based correlation or pattern for the filtered DEBIE-1 impacts. 
In this case the MASTER model shows a deviation in estimating the population by a factor of 2, but it models in concordance to the DEBIE-1 data in the sense that the overall trends match as shown in Figure \ref{fig:Cum_impacts}. The difference drifts further after 2016, which is during the forecast period for the MASTER model.

\section{Discussion}
The simulation tool used here, namely MASTER, is meant to model the general space particle environment and gives a smoothed flux over time by averaging the contribution of different events over a set of orbital parameters. 
A perfect match between detector and model is thus not to be expected, but methods like the developed cloud analysis with POEM can clarify how specific events alter the flux in the simulation tool.
Should a break-up occur from a collision or explosion resulting in a significant rise in flux on the detector, this approach allows the empirically fitting of the break-up parameters of the model.

Indeed, defining a volume that encompasses and tracks the evolution of a fragment cloud for  density determination would yield improved results in scenarios where the fragmentation event generated more sub-millimetre particles, or where the cloud evolves within a more compact space, i.e., where most fragments are contained within a limited volume over time.   
Such events where this would have been possible were however not observed by DEBIE-1. Nevertheless, the method outlined in this work demonstrates the essential considerations necessary for methods derived solely from simulation-based approaches such as \cite{TANAHASHI2024449}.

The analysis of DEBIE-1 focused on counting only fragments in the sub-millimetre regime, and hence favours detection of particles that are rather small which in this case coincides with SRM dust, ejecta, paint flakes and meteoroids. 
Based on the MASTER model assumptions, these are the sources that populate that size regime the most, and not necessarily space debris from fragmentation and explosion events. 
In this sense, recent trends to develop  smaller detectors (similar to DEBIE) as monitoring instruments capable of assessing the orbital surrounding for explosions and collisions might be of limited benefit.  
These events would need to be close enough to be detectable on a small device, distinct from the background particle environment, and simultaneously generate trackable fragments for ground based space surveillance systems.

Although speed and mass can be extracted from DEBIE-1 data, it has been reported in \cite{J_S_PhD} that due to the inherently noisy environment, the background plasma channel voltage varies throughout the orbit. The sampled voltage varies significantly even when a real impact has occurred, and an assumption of proportionality between the signal voltage and mass and speed does not hold.
As such, only the impact counts can reliably be used in this work. 
This also implies that impacts from natural and human-made particles cannot be differentiated from each other as speed is the only information that could possibly be used as a discriminating metric.

The search for the fragment clouds created by NOAA 16 and DMSP  5D-2 also raised a possible detection of meteoroid showers by DEBIE-1, that could in particular explain the high detections in summers 2015 and 2016 when the Perseids cloud peaks in activity, coinciding with the higher densities overall. 
Moreover, due to its imager instruments, PROBA-1 performs a rolling motion that could lead to an attitude bias of spacecraft enhancing meteoroid detection, as during the attitude changes DEBIE-1 spends less time facing the flight direction and more time oriented towards deep space.
In absence of a dedicated meteor stream model for these showers, no conclusions are drawn here.

The spacecraft has a second sensor, SU-2, on the starboard face, from which only a few tens of impacts were detected in total. 
With such a small sample size, no reliable statistical measurements of the flux could be established and the detected impacts were not included in this work.
This is a common shortcoming of a significant amount of in-situ experiments listed in Figure \ref{fig:in_situ_sensors}.

Currently only two space environment particle models are open and widely used in the community: ESA's MASTER and NASA's ORDEM.
While they exhibit strong agreement in predicting trackable space debris, up to two orders of magnitude difference arises in estimating particle fluxes smaller than 1 cm  \cite{horstmann2021flux}.
This divergence stems from distinct modelling approaches, data sources, and the limited availability of data in these size ranges. The impact detections filtered through DEBIE-1 SU-1, along with their derived flux levels, align with an environment that falls between the predictions of these two models. Hence, they serve as valuable additional calibration benchmarks for both models.

This study additionally highlights limitations of processing collected data from in-situ detectors. Hosting an in-situ sensor as an add-on should come with care as the optimal placement may be overlooked compared to the primary objectives of spacecraft missions. This leads to either excessive detected noise or suboptimal positioning caused by attitude and orientation changes. 
Even with optimal positioning, it is evident that these sensors would need to be in proximity in space and time to a fragmentation event to discern fragmentation clouds over the existing background population.  Differentiating between human-made and meteoroid populations would require additional measurements i.e., full velocity vectors to infer the origin of the particle. Sensors that can differentiate the material sources of the impactor are unfortunately few, complex and costly. Novel designs of in-situ sensors are now offering improved readings and directionality information as well as greater coverage which, when coupled with adequate data extraction and processing, can provide promising datasets to enhance the modelling of the unknown micro-world in space.

\section{Methods}
\subsection{Space debris and meteoroid environment model}\label{sec:Master}
For the analysis in this work, ESA's Meteoroid and Space Debris Terrestrial Environment Reference (MASTER) version 8.0.3 is extensively used \cite{MASTER}.
The purpose of the MASTER model is the realistic description of the natural and the human-made particulate environment in Earth orbit, enabling risk assessment via flux predictions on user-defined target orbits. 
The fluxes, defined as the amount of particles crossing a reference area per year, on user-defined target orbits are described down to impactor diameters of 1 {\textmu}m. 

Predictions for the historic space debris evolution from the start of the space era to a reference epoch of November 2016, and 30 years beyond that date. 
They are computed using models that simulate the generation of objects from all known debris sources, and their orbit evolution over time.
These sources include on-orbit fragmentations (explosions and collisions) which feature in this work as single events, solid rocket motor firings, coolant release from nuclear reactors in space, copper needles released in space, surface degradation due the extreme environment conditions in space, ejecta due to the impacts of small particles, and natural meteoroids including the time dependent streams.

Although the theoretical background of the models used for the generation of the sources is state-of-the-art, 
there still remains a high level of uncertainty for most of them. The population resulting from the simulation process therefore has to undergo a calibration against measurement data \cite{MASTER}.    
Those simulated space debris populations are subsequently validated by correlating with ground-based measurements for larger objects or through the analysis of impact features in returned surfaces for the more numerous non-observable part of the population (small object validation). 
This means that the model parameters responsible for generating small fragments are iteratively tuned for all sources and across all orbits until a good match with the samples is achieved. 
Population snapshots can then be validated against the historic population and result in the current reference population at 1 November 2016 \cite{VitaliModelsIAC}.

For computational efficiency, the number of parameters defining fluxes within the MASTER model is restricted. Specifically, fast moving Keplerian elements like true anomaly and RAAN are averaged over, making it challenging to directly discern the impact of a single event on short timescales \cite{MASTER}.

\subsection{Number of impacts estimation}\label{sec:midas}
To obtain a rough estimate of the expected number of impacts on any given surface in Earth orbit, a first order approximation can be derived from an analysis of collision flux.
The order of magnitude analysis, or the derived impact rates across a specified duration, can then be compared to the sensor detections. This helps identify periods that deviate from the anticipated average environmental behavior.
The MASTER-based Impact Flux and Damage Assessment Software (MIDAS) is based on the European space debris model MASTER, from which one can retrieve space debris and meteoroid collision flux and damage analysis for any user-defined target orbit and for particle size down to 1 {\textmu}m \cite{BRAUN2020206}.
For the collision flux analysis, a reference area $A$ ($\text{m}^2$) is used to sample the space environment in any orientation, for a given simulation time $\Delta t$ ($year$), yielding the impact flux $F$ ($1/\text{m}^2/year$).
The software can then compute the number of impacts as $FA\Delta t$.

\subsection{Fragment cloud modelling}\label{sec:poem}
\begin{figure}
\centering
\includegraphics[scale=0.6]{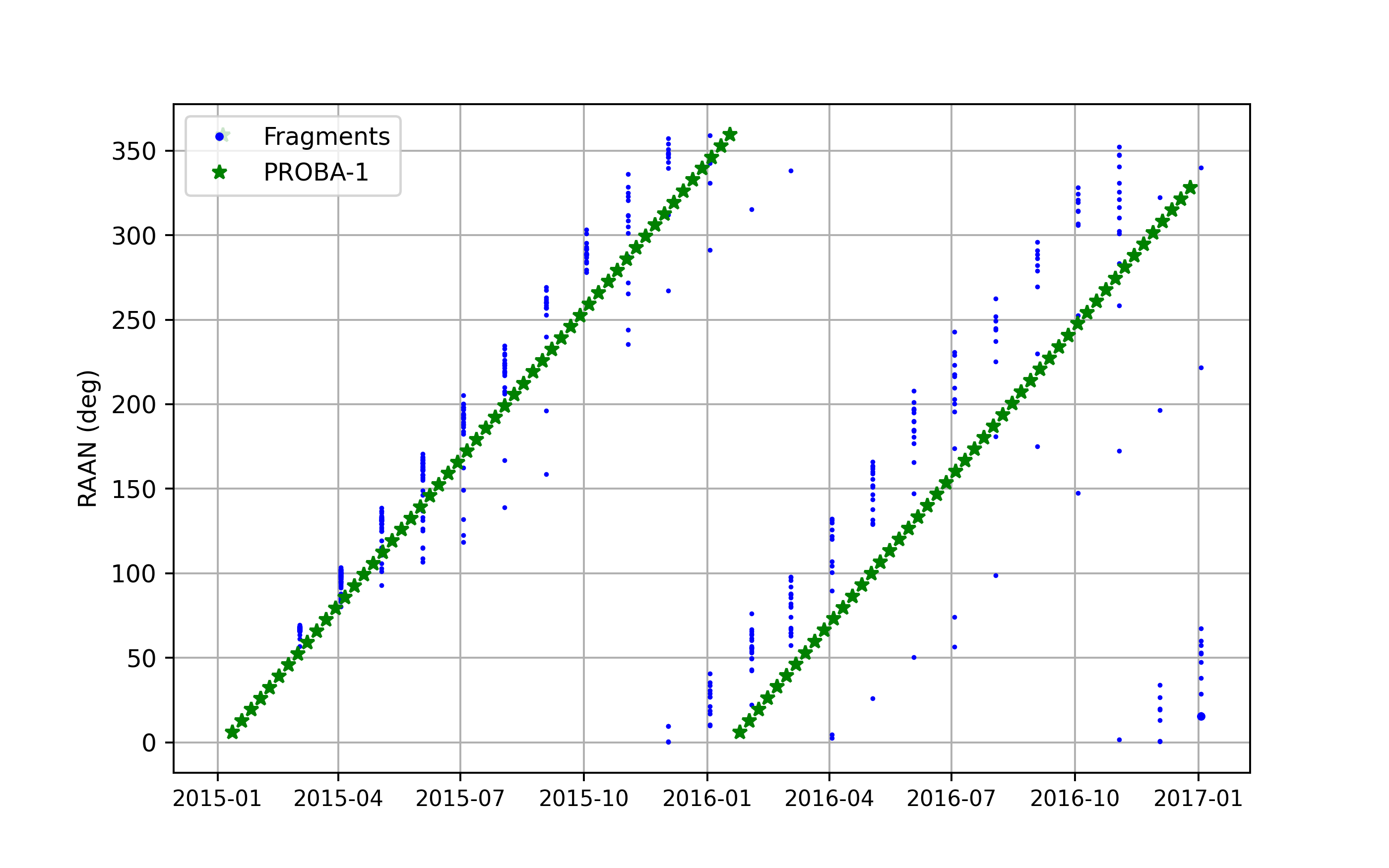}
\caption{RAAN evolution of PROBA-1 and the simulated fragments. The fragments are grouped in classes of similar size and orbit. Each blue marker indicates a fragment class. }
\label{fig:RAAN_evolution}
\end{figure}

Part of the generation of the MASTER model data is from a software suite called Program for Orbital Debris Environment Modelling (POEM) that simulates the generation and propagation of orbital debris objects larger than 1 {\textmu}m.
One of the space debris generation mechanisms implemented in POEM is the NASA breakup model for collision and explosive events \cite{johnson2001nasa}, with the event scaling mechanism of the original model replaced by a different calibration method derived from the existing Battelle model \cite{MASTER}. 
Each explosion and collision event in the environment is analysed to obtain an adequate scaling parameter between observed fragments and those predicted by the model. 
Space debris generated in such events are propagated from the event epoch to the common reference epoch by means of mean (singly averaged over mean anomaly) Kepler orbital elements integration of the averaged perturbation equations. 
The perturbations considered are zonal geopotential perturbations (J2 to J5), air drag (oblate atmosphere with diurnal density variations), 3rd body solar and lunar perturbations (up to 3rd order), and solar radiation pressure (with cylindrical shadow).
The propagator considers representative fragments for the size distribution created in the explosion or the collision model, and tracks these fragments over time. 
Figure \ref{fig:RAAN_evolution} illustrates the DMSP 5D-2 explosion modelling and the RAAN evolution of the generated fragment clouds over time. Theoretically, an event becomes detectable if the RAAN and altitude of the fragment cloud coincide with the orbit of PROBA-1.

\subsection{Density computation}\label{density definition} 

The debris density is used in this study as a metric to compare the detection by DEBIE-1 and the state-of-the-art model POEM. 
Analysing the simulated cloud and its associated orbits is essential to determine a potential detection by DEBIE-1. This involves examining eccentricity, perigee radius $r_p$, apogee radius $r_a$, and characteristics of the simulated fragments such as their estimated diameters.

POEM generates monthly clouds of fragments, only those in the $[10^{-6},10^{-3}]\,\text{m}$ size regime are considered, in accordance with the detection capabilities of DEDIE-1. 

Let $a$ be PROBA-1 orbit semi-major axis. From the fragment cloud simulated by POEM, we eliminate fragments with  $r_p \geq a+s$ and  $r_a\leq a-s$, where $s$ is a buffer altitude, an additional margin around PROBA-1. This encompasses either fragments with similar orbit as PROBA-1 or fragments with highly eccentric orbits that can still cross the orbit of PROBA-1.

\begin{figure}
    \centering
    \input{sphere.tex}
    \caption{One side of the double wedge volume considered to compute fragments density, delimited by an altitude interval centered around PROBA-1's semi-major axis, and by a RAAN interval enclosing most of the fragments.}
    \label{fig:wedge_visual}
\end{figure}
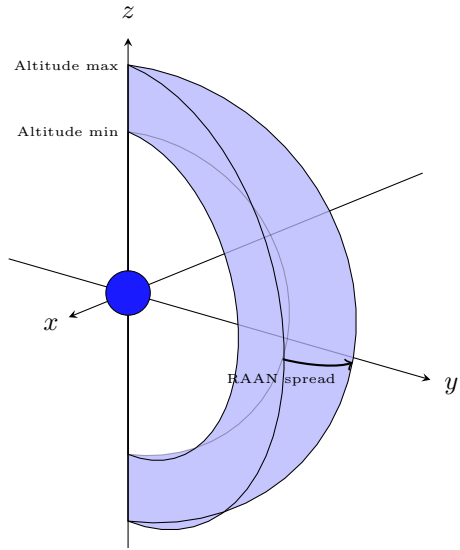

The filtering process provides a monthly catalogue of fragments that are in the vicinity of PROBA-1. The filtered fragments exist in a volume confined by an altitude interval around PROBA-1 semi-major axis, $[a-s,a+s]$, and further delimited by the evolving RAAN of the cloud over time, as depicted in Figure \ref{fig:wedge_visual}. 

This volume is enclosed between a larger sphere with a radius of $a+s$ and a smaller sphere with a radius of $a-s$. This enclosed space is then scaled by the angle representing the proportion of the spherical wedge, indicated as $\frac{\Delta \Omega}{2\pi}$, where $\Delta \Omega$ denotes the RAAN span of the fragments. The final volume encompasses two analogous wedges, hence the multiplication by $2$. Consequently, the density can be expressed as follows:

\begin{equation}
\rho_{POEM}(t) =\frac{\sum_t fragments }{2(\frac{\Delta \Omega}{2\pi})(4/3)\pi((a+s)^3-(a-s)^3)}
\end{equation}

Likewise, the volume monitored by DEBIE-1 is a surface of $10\times 10$ cm$^2$ rotating on a sphere with a radius of $a$. The density follows accordingly:

\begin{equation}
\rho_{DEBIE}(t) =\frac{\sum_t impacts }{0.1 \times 0.1 \times 2\pi a}
\end{equation}

\section{Data availability}\label{sec:data_availability}
The scientific data of DEBIE-1 has been retrieved from the European Detector Impact Database (EDID), hosted by ESA's Space Weather Services (ESA-SWE) at https://swe.ssa.esa.int/edid1 \cite{EDID}. 
A registered user can access the scientific data, the corresponding sensor uptimes and housekeeping data associated with the sensors. 
It is important to point out that the currently integrated `Real Impact' filters are only applicable to datasets up to mid-2005. 
For data beyond this date, one should download the raw datasets and post-process it using appropriate filters. 

The in-situ detector filtering algorithm developed in \cite{DMF04} has been used to process all raw data of DEBIE-1.
The algorithm supports, on top of the baseline channel filters developed and calibrated during the first years of operations \cite{J_S_PhD}, two dedicated noise reduction techniques. 
The first one is the terminator line filter, which removes noise events associated with thermal creak upon entering sunlight and possible spacecraft charging, as well as battery voltage peaks when entering Earth shadow. 
The second filter addresses passes above the Russian Kamchatka peninsula, which can reduce noise events on SU-2 caused by a radar on Shemya Island.  
This additional processing supplements an initial filtering stage that identifies 99 percent of impacts as noise \cite{bunte2021data}.

It is noted that during most of 2017 and up to the beginning of 2018 DEBIE-1 was unexpectedly picking up a large amount of impacts, even when taking into account the above mentioned filters. 
These erratic sensor readings are possibly due to a change in the instrument operation schedule due to prioritisation of other PROBA-1 payloads.
A credible correlation with on-orbit events could not be established other than the aforementioned change in operations of  spacecraft.
The months with erratic impact numbers were thus excluded from the study, as they all fall above the $93\%$ quantile of all detections between 2002 and 2022, and are considered outliers. 

The PROBA-1 orbits and attitude information files were obtained directly from the housekeeping database following communication with  spacecraft operations managers. These can be provided upon a reasonable request.

The historical data on past fragmentation events (event epochs and orbits) were retrieved from ESA's DISCOS (Database and Information System Characterising Objects in Space) database. DISCOS  serves as a single-source reference for launch information, object registration details, launch vehicle descriptions, spacecraft information (e.g. size, mass, shape, mission objectives, owner) as well as orbital data archives for all trackable, unclassified objects \cite{DISCOS}. 

\section{Code availability}
The Python-based software used to analyse impact detection data and compare it to the MASTER population is available on reasonable request.

\bibliographystyle{acm}
\bibliography{references}
\end{document}

%% file: sphere.tex
\begin{tikzpicture}[scale= 0.56]

\begin{axis}
[axis equal,axis lines = center,width = 20cm,height = 20cm,xlabel = {$x$},ylabel = {$y$},zlabel = {$z$},view/h=130,view/v=20,
    every axis x label/.style={ at={(ticklabel* cs:1.05)}},
    every axis y label/.style={ at={(ticklabel* cs:1.05)}},
    every axis z label/.style={ at={(ticklabel* cs:1.05)}},
    xtick=\empty,ytick=\empty,ztick=\empty,
    xmax=1.7,zmax=5,zmin=-5,]
\pgfmathsetmacro{\thetaStart}{80}
\xdef\thetaEnd{110}
\pgfmathsetmacro{\r}{sqrt(10)}
\pgfmathsetmacro{\rg}{sqrt(20)}

\node [left,font=\tiny] at (0,0,\r) {Altitude min};
\node [left,font=\tiny] at (0,0,\rg) {Altitude max};
\node [left, font=\tiny,thick] at (1, \rg +1.6,0){RAAN spread};

\addplot3[samples=30,opacity= 0.3, samples y=0,  domain=-90:90, name path=arc1] ({\r*cos(\thetaEnd)*cos(x)}, {\r*sin(\thetaEnd)*cos(x)}, {\r*sin(x)}) \closedcycle;
\addplot3[samples=30, samples y=0,  domain=-90:90, y domain = 2:3, name path=arc2] ({\rg*cos(\thetaEnd)*cos(x)}, {\rg*sin(\thetaEnd)*cos(x)}, {\rg*sin(x)}) \closedcycle;

\draw[->, thick] (0.9,\rg,0) arc [start angle=5,end angle=90,x radius=0.7,y radius=1.2];

\addplot3[samples=30, samples y=0,  domain=-90:90, name path=arc3] ({\r*cos(\thetaStart)*cos(x)}, {\r*sin(\thetaStart)*cos(x)}, {\r*sin(x)}) \closedcycle;
\addplot3[samples=30, samples y=0,  domain=-90:90, name path=arc4] ({\rg*cos(\thetaStart)*cos(x)}, {\rg*sin(\thetaStart)*cos(x)}, {\rg*sin(x)}) \closedcycle;

\addplot[blue!30, fill opacity=0.5] fill between [of=arc3 and arc4];
\addplot[blue!30, fill opacity=0.5] fill between [of=arc1 and arc2];
\addplot[blue!30, fill opacity=0.5] fill between [of=arc2 and arc4];
\addplot[blue!30, fill opacity=0.5] fill between [of=arc1 and arc3];

\filldraw [fill=blue!90] (0,0,0) circle (15pt);

\end{axis}
\end{tikzpicture}